\documentclass[onecolumn,showpacs]{revtex4}

\usepackage{graphicx}
\usepackage{dcolumn}
\usepackage{bm}

\input epsf

\begin{document}

\title{\Large Thermodynamics of Modified Chaplygin Gas and Tachyonic Field}

\author{\bf Samarpita Bhattacharya\footnote{samarpita$_{-}$sarbajna@yahoo.co.in}
and Ujjal Debnath\footnote{ujjaldebnath@yahoo.com ,
ujjal@iucaa.ernet.in}}

\affiliation{Department of Mathematics, Bengal Engineering and
Science University, Shibpur, Howrah-711 103, India.}

\date{\today}

\begin{abstract}
Here we generalize the results of the work of ref. [10] in
modified Chaplygin gas model and tachyonic field model. Here we
have studied the thermodynamical behaviour and the equation of
state in terms of volume and temperature for both models. We have
used the solution and the corresponding equation of state of our
previous work [12] for tachyonic field model. We have also studied
the thermodynamical stability using thermal equation of state for
the tachyonic field model and have shown that there is no critical
points during thermodynamical expansion. The determination of
$T_{*}$ due to expansion for the tachyonic field have been
discussed by assuming some initial conditions. Here, the thermal
quantities have been investigated using some reduced parameters.
\end{abstract}


\maketitle

\section{\normalsize\bf{Introduction}}

Data [1,2] collecting from type Ia Supernovae explosion suggest
that a new kind of matter with positive energy density and with
negative pressure is one of the main candidate behind the
acceleration of the present day universe. This new kind of matter
is commonly known as Q-matter or dark energy and very recently
WMAP data [3] have revealed that Q-matter possesses $ 70\% $ of
the universe. Cosmologists suggest many candidates for the dark
energy. Among these the first one is cosmological constant
$\Lambda$ [4]. But due to low energy scale than the normal scale
for constant $\Lambda$, the dynamical $\Lambda$ was introduced
which is known as quintessence [5]. Again at very early stage of
universe the energy scale for varying $\Lambda$ is not sufficient.
So to avoid this problem, known as cosmic coincidence [6], a new
field, called tracker field [7] was prescribed. In similar way
there are many models [8] in Einstein gravity to best fit the
data. At  the present epoch, a  lot of works has been done to
solve this quintessence problem and most popular candidates for
Q-matter has  so  far been a scalar field having a  potential
which generates  a sufficient negative pressure. Various dark
energy candidates have been proposed such as quintessence
mentioned, k-essence, tachyon, phantom, Chaplygin gas, holographic
dark energy etc [9].\\

Here we generalize the results of the work of Myung [10] in
modified Chaplygin gas model [11] and tachyonic field model [12].
Here we have studied the thermodynamical behaviour and the
equation of state in terms of volume and temperature for both
models. We have used the solution and the corresponding equation
of state of our previous work [12] for tachyonic field model.
Recently Santos et al [13] have studied the thermodynamical
stability in generalized and modified Chaplygin gas model. We'll
study the thermodynamical stability [13] using adiabatic and
thermal equation of states for the tachyonic field model [12] and
investigate the behaviour of critical points during
thermodynamical expansion.\\

\section{\normalsize\bf{Study of Thermodynamics}}

The Einstein field equations for homogeneous, isotropic and flat
FRW universe are given by

\begin{equation}
{H}^{2}= \frac{1}{3}\rho
\end{equation}

and

\begin{equation}
\dot{H}=-\frac{1}{2}(\rho + p)
\end{equation}

where $H(=\frac{\dot{a}}{a}$) is the Hubble parameter. The
continuity equation is given by

\begin{equation}
\dot{\rho} + 3 H (\rho + p) =0
\end{equation}

where $p$ is the isotropic pressure and $\rho$ is the energy
density of the fluid defined by

\begin{equation}
\rho=\frac{U}{V}
\end{equation}

Here, $U$ is the internal energy and $V$ is the volume of the
universe.\\

We apply the combination of first law and second law of
thermodynamics to the system with volume $V$. Then it leads to
[10]

\begin{equation}
TdS=d(\rho V)+ pdV = d((\rho +p)V)-Vdp
\end{equation}

The integrability condition of thermodynamic system is given by

\begin{equation}
\frac{\partial^{2}S}{\partial T \partial V} = \frac{\partial^{2}S
}
{\partial V \partial T }
\end{equation}

which leads to the relation between pressure, energy density and
temperature as

 \begin{equation}
dp= \frac{\rho + p}{T} dT
\end{equation}

From (5) and (7), we get

\begin{equation}
dS=d\left(\frac{( \rho + p )V}{T}\right)
\end{equation}

Therefore integrating, we get the expression of the entropy as

\begin{equation}
S=\frac{(\rho+p) V}{T}
\end{equation}

or, the temperature can be written as

\begin{equation}
T=\frac{(\rho + p)V}{S}
\end{equation}

In the next two sections, we consider two dark energy models: (i)
Modified Chaplygin gas and (ii) Tachyonic field and we study the
thermodynamical behaviours for these models.\\

\section{\normalsize\bf{MODEL I : Modified Chaplygin Gas}}

We start with an exotic fluid, named Modified Chaplygin Gas which
obeys the following adiabatic equation of state [11]

\begin{equation}
p=A\rho - \frac{B}{\rho^{\alpha}}
\end{equation}

and $A>0$, $B>0$ and $0\leq \alpha \leq1$.\\

Solving the equations (3) and (11), we get the expression of the
density as

\begin{equation}
{\rho} (a)= \left[ \frac{B}{1 + A} + \frac{C}{a^{3 (1 + A)(1 +
\alpha}}\right]^{\frac{1}{1 + \alpha}}
\end{equation}

where $C(>0)$ is a constant of integration.\\

Since $V$ is the volume, so we have $V = a^{3}$ and from equations
(12) and (11), we get the expressions of the density and the
pressure in terms of volume $V$ as

\begin{equation}
\rho ( V )= \left(\frac{B}{1 + A} + \frac{C}{V^{(1 + A) (1 +
\alpha)}}\right)^{\frac{1}{1 + \alpha}}
\end{equation}

and

\begin{equation}
p( V )= \frac{     \frac{A C}{V^{(1 + \alpha) (1 + A)}} -
\frac{B}{1 + A }}  {\left(\frac{B}{1 + A} + \frac{C}{V^{( 1 +
\alpha)(1 + A)}}\right)^{\frac{\alpha}{1 + \alpha}}   }
\end{equation}

Furthermore, the equation of the state $w( V )$ for modified
Chaplygin gas is given by

\begin{equation}
w( V )= \frac{p}{\rho}= \frac{\frac{AC}{V^{(1 +\alpha)(1 + A)}} -
\frac{B}{1 + A}}{\frac{B}{1 + A} + \frac{C}{V^{( 1 + \alpha)(1 +
A)}}}
\end{equation}

and the square speed of sound is given by

\begin{equation}
\upsilon^{2}\equiv \frac{\partial p}{\partial\rho}= A -
\frac{B}{\alpha}\left(\frac{B}{1 + A} + \frac{C}{V^{(1 + \alpha)(1
+ A}}\right)^{\frac{1 - \alpha}{1 + \alpha}}
\end{equation}

and consequently, considering equations (10), (13) and (14) we get
the expression of temperature as

\begin{equation}
T( V )= \frac{(1 + A) C}  {S V^{A + (1 + A)\alpha}}
\left(\frac{B}{1 + A} + \frac{C}{V^{( 1 + \alpha)(1 + A)}}
 \right)^{- \frac{\alpha}{1 + \alpha}}
\end{equation}

Since volume $V$ lies between 0 and $\infty$. So we see that the
temperature $T\rightarrow \infty$ as $V\rightarrow 0$ and
$T\rightarrow 0$ as $V\rightarrow\infty$, so the third law of
thermodynamics is satisfied for modified Chaplygin gas model.\\

From the integrability condition and using (7) and (11), we have

\begin{equation}
\frac{A\rho^{1 + \alpha} + B \alpha}{(A + 1) \rho^{2 + \alpha} -
B \rho} d \rho= \frac{dT}{T}
\end{equation}

Solving we get,

\begin{equation}
\left(\rho^{1 + \alpha} - \frac{B}{1 + A}\right)^{\frac{A +
\alpha( 1 + A)}{( 1 + A)(1 + \alpha)}} \rho^{-\alpha}=
\frac{T}{T_{*}}\left(\frac{B}{A}\right)^{\frac{A}{(1 + A) (1 +
\alpha)}}\left(\frac{1}{1 + A}\right)^{\frac{A + \alpha (1 +
A)}{(1 + A) (1 + \alpha)}}
\end{equation}

which is a relation between $\rho$ and $T$ and $T_{*}$ is the
integration constant.\\

Again, using $w=\frac{p}{\rho}$ and (11), the equation (19) gives

\begin{equation}
\left(\frac{B}{A - w} - \frac{B}{1 + A}\right)^{\frac{A + \alpha(
1 + A)}{( 1 + A)(1 + \alpha)}} \left(\frac{B}{A - w}\right)^{-
\frac{\alpha}{1 + \alpha}}=
\frac{T}{T_{*}}\left(\frac{B}{A}\right)^{\frac{A}{(1 + A) (1 +
\alpha)}}\left(\frac{1}{1 + A}\right)^{\frac{A + \alpha (1 +
A)}{(1 + A) (1 + \alpha)}}
\end{equation}

which is a relation between $w$ and $T$ provided $w\ne A$.\\

Finally the heat capacity can be calculated in terms of $V$ and
$T$ Tas

\begin{equation}
C_{V}(T)= V \frac{\partial \rho}{\partial T}= \frac{V}{T\rho}
\left(\frac{A \rho^{1 + \alpha} + B \alpha}{(A + 1) \rho^{1 +
\alpha} -
B}\right)=\frac{V}{T}\left(\frac{A-w}{B}\right)^{\frac{1}{1+\alpha}}\left(\frac{(1+\alpha)A-\alpha
w }{1+w} \right)
\end{equation}

where, $w$ is related with temperature $T$ governed by the equation (20).\\

For simplicity, we take $\alpha=1$ in the equation of state (11),
i.e.,

\begin{equation}
p=A\rho - \frac{B}{\rho}
\end{equation}

Solving the quadratic equation in $\rho$ and inserting the value
in (7) we get a quadratic equation of $p$ which is

\begin{equation}
(1 + A) p^{2} - \frac{ST}{V}(1 + 2 A) p + \left(\frac{A S^{2}
T^{2}}{V^{2}} - B\right)=0
\end{equation}

Solving the above equation and considering only the negative sign,
we obtain

\begin{equation}
p V= \frac{S T (1 + 2 A) - \sqrt{S^{2} T^{2} + 4 B(1 + A)
V^{2}}}{2 (1 + A)}
\end{equation}

For adiabatic process, we know that the entropy $S$ = constant. So
in the case, taking $S=2$, (24) gives the simpler form

\begin{equation}
p V= \frac{T (1 + 2 A) - \sqrt{T^{2} + B (1 + A) V^{2}}}{(1 + A)}
\end{equation}

On the other hand, we have the quadratic equation for energy
density $\rho$ (taking $\alpha=1$)

\begin{equation}
(1 + A) V \rho^{2} - S T \rho - B V= 0
\end{equation}

whose solution leads to

\begin{equation}
\rho V=\frac{S T}{2 (1 + A)} + \frac{1}{2 (1 + A)}\sqrt{S^{2}
T^{2} + 4 B (1+A)V^{2}}
\end{equation}

For the case $S=2$, it also reduces to the simpler form

\begin{equation}
\rho V=\frac{1}{1 + A} \left(T + \sqrt{T^{2} +  B  (1 + A) V^{2
}}\right)
\end{equation}

Now dividing (24) by (27) leads to the equation of state for
modified Chaplygin gas as functions of $T$ and $V$:

\begin{equation}
\omega(T,V)= A - \frac{B}{\rho^{2}(T,V)}
\end{equation}

Now from equations (24) and (27), we get\\

(i) $\rho\rightarrow \frac{ST}{(1+A)V}\approx\infty$ and
$p\rightarrow \frac{AST}{(1+A)V}=A\rho\approx\infty$ as $V\rightarrow 0$ provided $A\ne 0$.\\

(ii)  $\rho\rightarrow \sqrt{\frac{B}{1+A}}$ and $p\rightarrow
-\sqrt{\frac{B}{1+A}}=-\rho$ as $V\rightarrow\infty$ i.e., the
fluid becomes a cosmological constant.\\

\section{\normalsize\bf{Model II : Tachyonic field}}

The energy density $\rho$ and the pressure $p$ of the tachyonic
field are [12],
\begin{equation}
\rho=\frac{V(\phi)}{\sqrt{1-{\dot{\phi}}^{2}}}
\end{equation}
and
\begin{equation}
p=-V(\phi) \sqrt{1-{\dot{\phi}}^{2}}
\end{equation}

where $\phi$ is the tachyonic field and $V(\phi)$ is the
corresponding potential. Now take a simple form of
$V=\left(1-\dot{\phi}^{2}\right)^{-m}, ~(m>0)$ as described in the
ref. [12], so the solution of $V$ becomes

\begin{equation}
V=\left[1+\left(\frac{V^{*}}{a^{3}}\right)^{\frac{2}{1+2m}}
\right]^{m}
\end{equation}

where $V^{*}$ is a positive constant. Now using equations (1), (2)
and (30-32), we obtain the adiabatic equation of the state for the
tachyonic field (due to this particular solution) as

\begin{equation}
p= - \rho^{\frac{2 m - 1}{2 m + 1}}
\end{equation}

We derive the energy density equation from (30) as

\begin{equation}
\rho(a)=
\left[1+\left(\frac{V^{*}}{a^{3}}\right)^{\frac{2}{1+2m}}\right]^{\frac{2
m + 1}{2}}
\end{equation}

\subsection{\normalsize\bf{Adiabatic equation of state for tachyonic field}}

Now from equations (33) and (34), we obtain the energy density and
pressure in terms of volume $V$ as

\begin{equation}
\rho=\left[ 1 + \left({\frac{V^{*}}{V}}\right)^{\frac{2}{2 m +
1}}\right]^{\frac{2 m + 1}{2}}
\end{equation}

and

\begin{equation}
p= - \left[ 1 + \left({\frac{V^{*}}{V}}\right)^{\frac{2}{2 m +
1}}\right]^{\frac{2 m - 1}{2}}
\end{equation}

For small $V$, we find $\rho\approx\frac{V^{*}}{V}$ and $p\approx
- \left(\frac{V^{*}}{V}\right)^{\frac{2 m - 1}{2 m + 1}}$ which is
negatively very large  if $m>\frac{1}{2}$ and
 nearly equal to $0$ if $m<\frac{1}{2}$. Also
for large $V$, we see that $\rho\approx 1 + \frac{2 m + 1}{2}
\left(\frac{V^{*}}{V}\right)^{\frac{2}{2 m + 1}}$ and $p\approx -1
-\frac{2 m - 1}{2}
\left(\frac{V^{*}}{V}\right)^{\frac{2}{2 m - 1}}$.\\

Furthermore the equation of state $w(V)=p/\rho$ and square speed
of sound $v^{2}(V)$ are given by

\begin{equation}
w(V)= -\frac{1}{1 + \left({\frac{V^{*}}{V}}\right)^{\frac{2}{2 m +
1}}}
\end{equation}

and

\begin{equation}
v^{2}(V)= -\left(\frac{2 m-1}{ 2 m+1}\right)\frac{1}{1 +
\left({\frac{V^{*}}{V}}\right)^{\frac{2}{2 m + 1}}}=\left(\frac{2
m-1}{ 2 m+1}\right)~w(V)
\end{equation}

From Equation (36), we have
\begin{equation}
\left(\frac{\partial p }{\partial V}\right)_{S}= - \left(\frac{2 m
- 1}{2m +
1}\right)\frac{p}{V}\left({\frac{V^{*}}{V}}\right)^{\frac{2}{2 m +
1}}
\end{equation}

Since Tachyonic field propagated between dust and $\Lambda$CDM, so
pressure $p$ must be negative and $m$ must be satisfied $0<
m<\frac{1}{2}$. Therefore from (39) we must have [14]

\begin{equation}
\left(\frac{\partial p }{\partial V}\right)_{S}< 0
\end{equation}

i.e., pressure is reduced through the adiabatic expansion. So the
tachyonic fluid along its evolution is thermodynamically stable.
But this condition is not enough for stability of thermodynamics.
It is also necessary to determine if the pressure reduces or
remains constant as the fluid expands at constant temperature $T$,
in the same region where eq. (39) is negative [14]. Thus, one must
also verify if

\begin{equation}
\left(\frac{\partial p }{\partial V}\right)_{T}\le 0
\end{equation}

and the thermal capacity of the constant volume [32],

\begin{equation}
C_{V}>0
\end{equation}

then the tachyonic fluid along its evolution is fully
thermodynamically stable. These will be discussed in the following
subsection.\\

\subsection{\normalsize\bf{Thermal equation of state for tachyonic field}}

The equation of state for the tachyonic field is considered in
equation (33) and the energy density for tachyonic field is
considered in equation (4). From general thermodynamics, we write
[13, 14]

\begin{equation}
\left(\frac{\partial U }{\partial V}\right)_{S} = - p
\end{equation}

From (4), (33) and (43), we obtain

\begin{equation}
\left(\frac{\partial U }{\partial V}\right)_{S}
=\left({\frac{U}{V}}\right)^{\frac{2 m - 1}{2 m + 1}}
\end{equation}

whose solution is given by

\begin{equation}
U=V \left[ 1 + \left({\frac{V^{*}}{V}}\right)^{\frac{2}{2 m +
1}}\right]^{\frac{2 m + 1}{2}}
\end{equation}

Here, $V^{*}$ may be considered as a function of entropy $S$ or a
universal constant. From general thermodynamics [14], the
temperature $T$ of the tachyonic field can be determined from the
following equation:

\begin{equation}
T=\left(\frac{\partial U }{\partial
S}\right)_{V}=(V^{*})^{\frac{1-2m}{1+2m }}\left[V^{\frac{2}{2 m +
1 }} +(V^{*})^{\frac{2}{2 m + 1 }}\right]^{\frac{2 m - 1}{2}}
\frac{d V^{*}}{d S}
\end{equation}

If $V^{*}$ is chosen as a universal constant then $\frac{d
V^{*}}{d S}=0$. In such conditions, the temperature will be zero
for any value of the volume or pressure of the gas. Thus, the
isotherm $T = 0$ is simultaneously an isentropic curve (adiabatic)
at $S =$ constant and this violates the third law of
thermodynamics. Therefore, to discuss extensively the
thermodynamic stability of the tachyonic field, it is necessary to
assume that the condition
is not satisfied.\\

If $V\gg V^{*}$, that is V is very large, then (46) reduces to

\begin{equation}
T\approx\left(\frac{V}{V^{*}}\right)^{\frac{2 m - 1}{2 m + 1
}}\frac{d V^{*}}{d S}
\end{equation}

Since $T>0$ for adiabatic expansion, so from (47), we must have
$\frac{dV^{*}}{dS}>0$. Now from (45), we see that the dimension of
$V^{*}$ is same as the dimension of $U$. So, we may write $V^{*}$
as [13] $V^{*}=T_{*}S$, where $T_{*}$ is some constant related to
the temperature. So $T_{*}=\frac{dV^{*}}{dS}$. Now, from (46), we
obtain,

\begin{equation}
T=T_{*}^{\frac{2}{2 m + 1}}S^{\frac{1 - 2 m}{1 + 2
m}}\left[V^{\frac{2}{2 m + 1 }} + T_{*}^{\frac{2}{2 m + 1
}}S^{\frac{2}{1 + 2 m}}\right]^{\frac{2 m - 1}{2}}
\end{equation}

By solving (48), we obtain the expression of entropy $S$ as

\begin{equation}
S=\frac{V}{T_{*}}\left[\left(\frac{T}{T_{*}}\right)^{\frac{2}{2 m
- 1}} - 1\right]^{-\frac{2 m + 1}{2}}
\end{equation}

Since in our tachyonic field model, $m<\frac{1}{2}$ and entropy
$S$ is always positive. So from above result, we can conclude that
$0<T<T_{*}$.\\

From equations (7) and (33), we get

\begin{equation}
\frac{dp}{p-(-p)^{\frac{2m+1}{2m-1}}}=\frac{dT}{T}
\end{equation}

and after integration we obtain the expression pressure in terms
of temperature as

\begin{equation}
p=-\left[1-\left(\frac{T}{T_{*}}\right)^{\frac{2}{1-2m}}\right]^{\frac{1-2m}{2}}
\end{equation}

and using eq. (33), we obtain the expression for density as

\begin{equation}
\rho=\left[1-\left(\frac{T}{T_{*}}\right)^{\frac{2}{1-2m}}\right]^{-\frac{2m+1}{2}}
\end{equation}

 and hence the equation of state is obtained as

\begin{equation}
w(T)=\frac{p}{\rho}=-1+\left(
\frac{T}{T_{*}}\right)^{\frac{2}{1-2m}}
\end{equation}

From the above expression, we see that $\rho\rightarrow 1$ and
$p\rightarrow -1$ as $T\rightarrow 0$ and $\rho\rightarrow\infty$
and $p\rightarrow 0$ as $T\rightarrow T_{*}$ for $m<\frac{1}{2}$
which is expected in our model. So $T_{*}$ defines the temperature
of dust and zero temperature occurs at $\Lambda$CDM stage.. We
also see that, $p$ depends only on $T$ for any volume $V$. So we
have $\left(\frac{\partial p }{\partial V}\right)_{T}= 0$. Thus
the second condition of the stability of thermodynamics is always
satisfied. It is also interesting that the all derivatives of $p$
with respect to $T$ become zero for any volume $V$. Thus
there is no critical points during the thermodynamical expansion.\\

Now the thermal capacity of the constant volume [14] can be
calculated as

\begin{equation}
C_{V}=T\left(\frac{\partial S }{\partial T}\right)_{V}=-\frac{2 m
+ 1}{2 m - 1}\frac{V}{
T_{*}}\left(\frac{T}{T_{*}}\right)^{\frac{2}{2 m - 1
}}\left[\left(\frac{T}{T_{*}}\right)^{\frac{2}{2 m - 1 }} -
1\right]^{-\frac{2m + 3}{2}}
\end{equation}

Since $0<m<\frac{1}{2}$ and $0<T<T_{*}$, so (54) shows that
$C_{V}>0$ in the tachyonic field model. So the third condition of
stability is always satisfied. So we may conclude that tachyonic
field has a thermodynamically stable behaviour
throughout the expansion.\\

\subsection{\normalsize\bf{Determination of temperature}}

Now we have to determine the temperature $T_{*}$ due to expansion
for the tachyonic field. For this this purpose, let us assume the
initial conditions: $\rho=\rho_{0},~p=p_{0},~V=V_{0}$ and
$T=T_{0}$. So from equations (33) and (35), we obtain

\begin{equation}
V^{*}=V_{0}\left(\rho_{0}^{\frac{2}{2 m + 1}} -
1\right)^{\frac{2m+1}{2}}
\end{equation}
and
\begin{equation}
p_{0}= - \rho_{0}^{\frac{2 m - 1}{2 m + 1}}
\end{equation}

Equations (35), (36) and (55), we obtain the energy density $\rho$
and pressure $p$ as a function of the volume $V$ :

\begin{equation}
\rho=\left[1 + \left(\rho_{0}^{\frac{2}{2m+1}}-1 \right)
\left(\frac{V_{0}}{V}\right)^{\frac{2}{2 m + 1}}\right]^{\frac{2 m
+ 1}{2}}
\end{equation}
and
\begin{equation}
p= -\left[1 + \left(\rho_{0}^{\frac{2}{2m+1}}-1 \right)
\left(\frac{V_{0}}{V}\right)^{\frac{2}{2 m +
1}}\right]^{\frac{2m-1}{2}}
\end{equation}

Now define some reduced parameters, like

\begin{equation}
\epsilon=\frac{\rho}{\rho_{0}},~\eta=\frac{p}{p_{0}},~\gamma=\frac{1}{\rho_{0}^{\frac{2}{2m+1}}},~v=\frac{V}{V_{0}},~
\tau=\frac{T}{T_{0}},~ \tau_{*}=\frac{T}{T_{*}}
\end{equation}

Equation (58) can be written as

\begin{equation}
p=-\left[1-\left(\frac{\tau}{\tau_{*}}\right)^{\frac{2}{1-2m}}\right]^{\frac{1-2m}{2}}
\end{equation}

Therefore, using (59), the equations (57) and (58) become,

\begin{equation}
\epsilon= \left[\gamma + \frac{1 - \gamma}{\upsilon^{\frac{2}{2 m
+ 1}}}\right]^{\frac{2 m + 1}{2}}
\end{equation}
and
\begin{equation}
\eta= \left[\gamma + \frac{1 - \gamma}{\upsilon^{\frac{2}{2 m +
1}}}\right]^{\frac{2 m - 1}{2}}
\end{equation}

Also from (59), (61) and (62), we see that at
$\rho=\rho_{0},~p=p_{0}$, $V=V_{0}$, $T=T_{0}$:

\begin{equation}
\epsilon=1,~\eta=1,~\tau=1,~ v=1,~  p_{0}=
-\gamma^{\frac{1-2m}{2}},~ \tau_{*}=\gamma^{\frac{1-2m}{2}}
\end{equation}

For example, the maximum temperature of the fluid as
$T_{*}=10^{32}$K, the temperature of the Planck era, and the
temperature of the tachyonic fluid at the present epoch, $T_{0}$ =
2.7K [13], then we obtain the value of $\gamma$ as $\gamma\approx
10^{\frac{64}{1-2m}}$. If we choose $m=1/3$,
we get $\gamma\approx 10^{192}$.\\

\section{\normalsize\bf{Discussions}}

In this work, we have considered the general thermodynamical
description for dark energy - modified Chaplygin gas and tachyonic
field models in flat FRW universe. The thermal quantities i.e.,
pressure, density etc. have been described in either functions of
volume or temperature. For adiabatic case with $\alpha=1$, the
thermal equation of state have been found and modified Chaplygin
gas cools down during expansion of the universe. For tachyonic
field model we have use the solution of our previous work [12] and
find the corresponding equation of state. For tachyonic field
model, we have used the adiabatic and thermal equation of state
[13]. For adiabatic equation of state, we find the thermal
quantities in terms of volume $V$ and also for thermal equation of
state, the thermal quantities have been found in terms of
temperature $T$. In both modified Chaplygin gas and tachyonic
field models, the dark energy temperature decreases from $T^{*}$
to zero for the expansion of the universe. The entropy $S$ has
been calculated in terms of $T$. It has been found that there is
no critical points for these dark energy models. For thermal
equation of state, we show that pressure $p$ depends on $T$ for
any volume $V$. The heat capacity and speed of sound have been
found in terms of volume $V$ and temperature $T$. Here, we have
shown that $\left(\frac{\partial p }{\partial V}\right)_{S}< 0$,
$\left(\frac{\partial p }{\partial V}\right)_{T}= 0$ and
$C_{V}>0$. So from these conditions, we conclude that tachyonic
field has a thermodynamically stable behaviour throughout the
expansion. The determination of $T_{*}$ due to expansion for the
tachyonic field have been discussed by assuming some initial
conditions. Here, the thermal quantities have been investigated
using some reduced parameters.\\

{\bf Acknowledgement:}\\

The authors are thankful to IUCAA, Pune, India for warm
hospitality where part of the work was carried out.\\

{\bf References:}\\
\\
$[1]$ S. Perlmutter et al, {\it Nature} {\bf 391} 51 (1998);
S. Perlmutter et al, {\it Astrophys. J.} {\bf 517} 565 (1999).\\
$[2]$ P. M. Garnavich et al, {\it Astrophys. J.} {\bf 493} L53
(1998); A. G. Riess et al, {\it Astron. J.} {\bf 116} 1009 (1998).\\
$[3]$  S. Bridle, O. Lahav, J. P. Ostriker and P. J. Steinhardt,
{\it Science} {\bf 299} 1532 (2003); C. Bennett et al, {\it
Astrophys. J. Suppl.} {\bf 148} 1 (2003); D. N. Spergel et al,
{\it Astrophys. J. Suppl.} {\bf 148} 175 (2003);\\
$[4]$  B. Ratra and P. J. E. Peebles, {\it Phys. Rev. D} {\bf 37}
3406 (1988).\\
$[5]$ R. R. Caldwell, R. Dave and P. J. Steinhardt, {\it
Phys. Rev. Lett.} {\bf 80} 1582 (1998).\\
$[6]$ P. J. Steinhardt, L. Wang  and I. Zlatev,{\it Phys. Rev. Lett.}, {\bf 59} 123504 (1999).\\
$[7]$ I. Zlatev, L. Wang and P. J. Steinhardt, {\it Phys. Rev.
Lett.} {\bf 82} 896 (1999).\\
$[8]$ V. Sahni and A. A. Starobinsky, {\it Int. J. Mod. Phys.}
{\bf 9} 373 (2003); T. Padmanabhan, {\it Phys. Rept.} {\bf 380}
235 (2003).\\
$[9]$ A. Sen, \textit{JHEP} \textbf{0207} 065 (2002); F. Piazza
and S. Tsujikawa, \textit{JCAP} \textbf{0407} 004 (2004); Z. K.
Guo, Y. S. Piao, X. M. Zhang and Y. Z. Zhang, \textit{Phys. Lett.
B} \textbf{608} 177 (2005); V. Sahni and Y. Shtanov, \textit{JCAP}
\textbf{0311} 014 (2003); A. Y. Kamenshchik, U. Moschella and V.
Pasquier, \textit{Phys. Lett. B} \textbf{511} 265
(2001).\\\
$[10]$ Y. S. Myung, arXiv:0812.0618v1 [gr-qc].\\
$[11]$ H. B. Benaoum, {\it hep-th}/0205140; U. Debnath, A.
Banerjee and S. Chakraborty, {\it Class. Quantum Grav.} {\bf 21}
5609 (2004); T. Bandyopadhyay and S. Chakraborty,
{\it Mod. Phys. Lett. A} {\bf 24} 2377 (2009).\\
$[12]$ S. Chattopadhyay, U. Debnath and G. Chattopadhyay, {\it
Astrophys. Space Sci.} {\bf 314} 41 (2008).\\
$[13]$ F.C. Santos, M. L. Bedran and V. Soares {\it Phys. Lett. B}
{\bf 636} 86 (2006); {\it Phys. Lett. B}
{\bf 646} 215.\\
$[14]$ L. D. Landau and E. M. Lifschitz, {\it Statistical
Physics}, third ed., Course of Theoretical Physics, Vol. 5,
Butterworth-Heinemann, London, 1984.\\

\end{document}